\documentclass[aps,twocolumn,amsmath,amssymb,superscriptaddress]{revtex4}

\usepackage[latin1]{inputenc}
\usepackage{graphics}
\usepackage{dcolumn}
\usepackage{graphicx}
\usepackage{dcolumn}
\usepackage{bm}

\providecommand{\qperpv}{\ensuremath{\bf{q} \! \perp \! \bf{v}}}
\providecommand{\qparav}{\ensuremath{\bf{q} \! \parallel \! \bf{v}}}

\begin{document}                  



\title{X-ray photon correlation spectroscopy under flow}

\author{Andrei Fluerasu}
\author{Abdellatif Moussa\"id}
\author{Henri Gleyzolle}
\affiliation{European Synchrotron Radiation Facility, Grenoble, France}
\author{P\'eter Falus}
\affiliation{Institut Laue-Langevin, Grenoble, France}
\author{Anders Madsen}
\affiliation{European Synchrotron Radiation Facility, Grenoble, France}

\date{\today}

\begin{abstract}
X-ray photon correlation spectroscopy was used to probe the diffusive dynamics  
of colloidal particles in a shear flow.
Combining X-ray techniques with microfluidics is an experimental strategy 
that  reduces the risk of x-ray induced beam damage and also allows 
time-resolved studies of processes taking place in flowcells.
The experimental results and theoretical predictions presented here, show 
that in the low shear limit, for a ``transverse flow'' scattering 
geometry (scattering wave vector ${\bf q}$ perpendicular to the direction 
of flow) 
the measured relaxation times are independent of the flow rate and 
determined only by the diffusive motion of the particles.
This is not generally valid and in particular, for a ``longitudinal flow''
(${\bf q}\parallel$~flow) scattering geometry, the relaxation times
are strongly affected by the flow-induced motion of the particles.
Our results show that the Brownian diffusion of colloidal particles 
can be measured in a flowing sample
and that, up to flux limitations, the experimental conditions under which 
this is possible are easier to achieve at higher values of $q$.
\end{abstract}

\maketitle                        


\section{Introduction}

Over the past decade, X-ray photon correlation spectroscopy 
(XPCS) has emerged as an unique experimental tool that allows the 
direct measurement of fluctuations in a large number of  condensed 
matter systems \cite{Mark_chapter}.
It provides a method complementary to Dynamic Light Scattering (DLS) 
\cite{Berne_Pecora} for the observation of mesoscale dynamics 
(e.g.~1~nm--1~$\mu$m length scales) in opaque materials 
or in samples where multiple scattering limits the applicability 
of DLS. 
Indeed, a problem more commonly encountered with X-rays, is a small 
scattering cross section, which means that multiple 
scattering can usually be neglected, but also that very intense 
beams are required to achieve reasonable signal to noise ratios. 
As a consequence, XPCS experiments have become possible only at 
high-brilliance Synchrotron Radiation (SR) sources.

In many recent applications, XPCS was used to study bulk 
equilibrium and/or non-equilibrium mesoscale dynamics in a 
large class of complex fluids including, but not being limited to, 
colloidal suspensions \cite{banchio_PRL06}, 
colloidal gels \cite{gel_PRE_07},
or polymer-based systems \cite{Falus_PRL97_2006}.
However, irradiation damage is often encountered when intense X-ray 
beams are employed to study soft-matter or biological samples. 
This problem will become even more important at the 
next generation of light
sources -- X-ray Free Electron Lasers (XFEL) and 
Energy-Recovery Linacs (ERL) -- with
their unprecedented brilliance, stronger by several orders of magnitude
than present third-generation SR sources 
\cite{Shenoy_FEL_03,Bilderback}. 
Flowing a fluid sample through a microfluidic device while
performing XPCS provides a method that can limit the beam 
induced damage effects and may allow the direct measurement of 
slow mesoscale dynamics in
various ``X-ray sensitive'' systems (e.g. colloids, polymers and 
bio-polymers, gels, aggregating proteins, etc.). 
In addition, 
this experimental strategy offers the possibility to perform 
time-resolved experiments. If a process like protein 
folding \cite{Pollack_PRL01} occurs in a microfluidic device, 
the time-dependence of the kinetics is mapped into a
spatial-dependence of stationary properties along the microfluidic channel.
In such an experimental configuration, XPCS could be used to study the 
time-dependence of dynamical properties even for very weekly scattering 
systems (e.g. formation of colloid and polymer gels, aggregating proteins). 

In the experiments reported here, the dynamics of a colloidal
suspension of hard-spheres in a shear flow was studied by XPCS. 
These results provide, to our knowledge, the first feasibility study
of XPCS as a probe for the diffusive dynamics in a flowing sample. 
In the visible range, similar experiments have been performed
by Ackerson and Clark using DLS \cite{Ackerson_Clark_JPhysique81}. 
Here we show that with the higher values of the scattering wave vector 
${\bf q}$ probed by X-rays, it is easier to achieve the experimental 
conditions under which the diffusive dynamics is accessible. 

XPCS employs a partially coherent X-ray beam which creates, 
when scattered from a disordered sample,
a characteristic \emph{speckle pattern} reflecting the 
instantaneous spatial arrangement of the scatterers \cite{Mark_Nature}.  
The technique consists in monitoring the temporal correlations of the 
speckle fluctuations, which are caused by the motion of 
the scatterers in the sample.

With a dilute colloidal suspension under shear flow, the correlation 
functions measured by XPCS are not only determined by the random 
(Brownian) motion of the colloids, but are also affected by their 
flow-induced motion.
As shown in the following, the correlation functions are strongly
influenced by the Doppler shifts 
resulting from particles moving at different average flow velocities due to 
the shear. The intensity scattered by particles 
moving with an average velocity difference of $\delta {\bf v}$, creates 
a signal on the detector that is modulated by
a self-beat frequency of ${\bf q}\cdot\delta {\bf v}$ \cite{Fuller}. 
Hence  a homodyne photon correlation spectroscopy experiment can measure 
velocity
gradients but not the absolute velocity, which is only accessible by 
heterodyne detection \cite{mark_heterodyne_07}.
It is also clear that, due to these shear-induced effects, the 
dynamics is not isotropic. The results detailed here show that
for a transverse flow geometry (scattering wave vector \qperpv),
the scalar products ${\bf q}\cdot\delta {\bf v}$ are all zero, 
and the relaxation times measured in a homodyne XPCS experiment  
are independent of the flow and measure only the diffusive dynamics of
the particles. This result is not generally valid, and in particular, 
the dynamics is strongly affected in the longitudinal flow (i.e. \qparav) 
direction. In this situation, the decay profile of the correlation 
functions are altered by the flow-induced distribution of
particle velocities,
and it is difficult, or impossible to ``extract'' their diffusive behavior. 

\section{Description of the experiments}

The sample, a suspension of sterically stabilized
polymethylmethacrylate (PMMA) particles dispersed in cis-decalin, is a
well characterized hard-sphere model system \cite{Abdellatif_colloid}.
Here, the radius of the PMMA particles was $a\approx$255 nm (with a size
polydispersity of 7\%), and the colloid volume fraction was $\Phi$=0.16 . The
flow cell was made out of a quartz capillary tube with a diameter of 
$\approx$980~$\mu$m (inset in Fig.~1a). The (relatively)
large diameter of the capillary was chosen to increase the sample volume
and the total scattering cross section which is often the limiting factor
for XPCS.  A syringe pump was 
pushing the fluid through $\approx$1mm Teflon tubes into the capillary.

\begin{figure}
\resizebox*{0.9\columnwidth}{!}{
\includegraphics{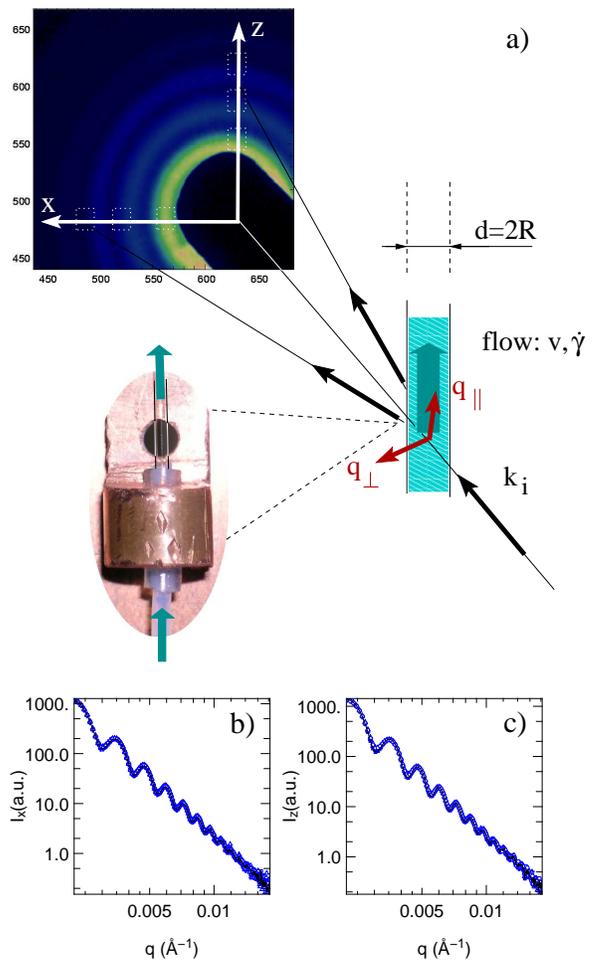}} 
\caption{\label{fig:flowstatics} (a) Small-angle X-ray experimental 
setup and flow chamber for the XPCS experiments. 
Static scattering, I$_x$, I$_z$, 
for the two flow 
geometries considered here - (b) transverse, \qperpv\ and
respectively - (c) longitudinal, \qparav. The data were
recorded at several flow rates ($v$=0-200 $\mu$m/s). 
In the $q$-range accessible here, the static 
properties are both isotropic and flow-independent. }
\end{figure}

Some important aspects about the fluid behavior, and in particular the laminar 
character of the flow are determined by the relative ratio of the inertial to 
viscous forces. This ratio is expressed as the \emph{Reynolds number} 
\cite{microfluidics},
\begin{equation}
\text{Re}=\frac{v d \rho}{\eta}
\label{eq:Reynolds}.
\end{equation}
Here $d$ is the characteristic length of the system (in our case the
capillary diameter $d = 2R$), $\rho$ is the fluid density, 
$v$ is the volume flow velocity (measured by the volume flow rate 
$Q=\pi R^2 v)$ of 
the fluid and $\eta$ is the dynamic viscosity. 
The onset of turbulent flow occurs at 
Reynolds numbers larger than $\approx$1000. In all the experiments reported 
here, the Reynolds numbers have values that are much smaller (Re~$<$~0.1) which
ensures a laminar flow. The main results of this paper, showing 
the (purely) diffusive nature of the dynamics measured in a transverse 
flow geometry confirm this conclusion. 

The exact velocity profile in the capillary is determined by the volume flow 
rate and the details about the boundary conditions at the capillary-fluid 
interface. In the results reported here, a simplified picture is adopted, 
and the flow is considered to be characterized by a single (average) shear 
rate $\dot\gamma=\frac{dv}{dx}$. Assuming a no slip boundary conditions model, 
the magnitude of this shear rate can be estimated by
\begin{equation}
\dot \gamma=\frac{3v}{R}.
\label{eq:shearrate}
\end{equation}

The XPCS experiments were performed in a small-angle x-ray scattering
geometry (Fig.~1) using partially
coherent X-rays at the ID10A beamline (Tro\"ika) of the European
Synchrotron Radiation Facility. A single bounce Si(111) crystal
monochromator was used to select 8~keV X-rays, having a relative
bandwidth of $\Delta \lambda /\lambda \approx  10^{-4}$. Higher order light
was suppressed by a Si mirror downstream of the monochromator,
and a transversely coherent beam was defined by a pinhole 
of diameter s=10~$\mu$m, placed 0.2~m upstream of the sample. 
The parasitic scattering from the 
pinhole was limited by a guard slit (corner)  placed in front of the sample. 
Under these conditions, the flux through the pinhole 
was of $\sim 10^9$~ph/s. The scattering from the
PMMA particles was recorded by a 0D scintillator detector (Cyberstar) located 
2.3~m downstream of the sample. The detector area was limited to a few 
speckle size (typically 50-100~$\mu$m) by precision slits in front of
the detector. Static data was also obtained using a CCD area-detector with
22.5~$\mu$m pixel size placed at the same distance (Fig.~1). 

The intensity autocorrelation functions,
\begin{equation}
	g^{(2)}({\bf q},t)=\frac{ \left< I({\bf q},0)I({\bf q},t) \right>}
	{\left< I({\bf q},t)\right>^2 },
\label{eq:g2}
\end{equation}
were obtained using a digital real-time hardware correlator (from 
correlator.com) 
connected to the X-ray detector. 
Assuming a Gaussian distribution of the temporal intensity fluctuations at a 
fixed ${\bf q}$, the intensity correlation functions are related to the 
dynamic structure factor or intermediate scattering function 
$f({\bf q}, t)$ via the Siegert relationship,
$
g^{(2)}({\bf q}, t)=1+\beta \left|f({\bf q},t)\right|^2.
$
Here $\beta$ is the speckle contrast, in this
setup around 5\% depending on the exact pinhole and detector 
slit sizes.

\section{XPCS in a laminar flow}

The correlation functions measured in a 
XPCS experiment on a fluid undergoing shear flow are 
determined by several factors:\\
{\bf i)} the (shear enhanced) diffusive motion of the particles;\\
{\bf ii)} the shear-induced distribution of average flow velocities, 
or more precisely the Doppler shifts coming from particles moving 
with different average velocities in the scattering volume;\\
{\bf iii)} the (average) transit time of the particles through the scattering volume. 

Each of these effects will be discussed in the following.

The diffusive motion of the colloidal particles is enhanced by the 
shear. This effect has been studied using DLS by Ackerson and Clark \cite{Ackerson_Clark_JPhysique81}.
The contribution of thermal diffusion to the intermediate scattering 
function of a colloidal suspension in a shear flow is shown to be 
described by
\begin{equation}
f_D({\bf q},t)=\exp\left[-\Gamma t\left(1 -\frac{q_\| q_\perp}{q^2} \dot\gamma t
	+ \frac{q_\| ^2}{q^2} \frac{(\dot\gamma t)^2}{3}\right)\right].
\label{eq:ISF_AckersonClark}
\end{equation}
Here ${\bf q}$ is the scattering wave vector with Cartesian components $q_\|$ and $q_\perp$ (parallel and respectively perpendicular to the direction
of the flow) and absolute value $q$, $\dot\gamma$ is the shear rate 
(considered uniform) and $\Gamma$ is the relaxation rate which relates
to $q$ and the diffusion constant $D_0$ via
\begin{equation*}
\Gamma =D_0 q^2.
\end{equation*}

In a transverse flow scattering geometry $q_\|$=0, hence 
Eq.~\ref{eq:ISF_AckersonClark} is independent of the flow (shear), and becomes
indistinguishable from the intermediate scattering function of a 
suspension undergoing a simple Brownian motion,
$f_{\perp} (q, t)=\exp\left(-D_0q^2 t \right)$.

The relevant time scale associated with thermal diffusion, the diffusion 
time, can thus be defined as
\begin{equation}
\tau _D=\frac{1}{\Gamma }=\frac{1}{D_0 q^2}.
\label{eq:diffusiontime}
\end{equation}

As it was shown in a number of \emph{Doppler velocimetry} experiments, 
\cite{Fuller,Narayan_AO97}, the intensity
correlation functions are not only determined by the diffusive motion of the
colloids (Eq.~\ref{eq:ISF_AckersonClark}), but are also modulated by a 
self-beat frequency created by 
particles moving with different average (flow) velocities. 
If the (shear-induced) velocity difference between two particles separated 
by a distance {\bf r}={\bf r}$_1$-{\bf r}$_2$ is $\delta${\bf v}, 
the resulting beating frequency is given by ${\bf q}\cdot\delta{\bf v}(r)$. 
This shear-induced effect can be detected only in ``non-transverse''
scattering geometries when the scalar product ${\bf q}\cdot\delta{\bf v}$
is different from zero.
The resulted intensity correlation 
function is thus modulated by an average
over all the Doppler shifts between all pair of particles in the
scattering volume, which can be written as
\begin{equation}
G_{\delta { v}} ({\bf q}, t)=\frac{1}{R^2} \int _0 ^R dr_1 \int _0 ^R dr_2 
	\exp \left( -i {\bf q}\cdot\delta {\bf v}(r) t \right),
\label{eq:Narayan_dv}
\end{equation}

In the case of a uniform shear, the double integral in Eq.~\ref{eq:Narayan_dv} 
can be calculated analytically \cite{Narayan_AO97}, leading to
\begin{equation}
G({\bf q},t)=\left[\frac{\sin \left(\Gamma _S\:t \right)}
	{\Gamma _S\:t}\right]^2
\label{eq:sinx_x},
\end{equation}
where the \emph{shear relaxation rate} $\Gamma_S$ depends on $q$ and the
flow velocity $v$ (or equivalently, on the flow rate $\dot\gamma$) and  
is given by $ \Gamma_S=q_{\|}v$.
A characteristic time scale associated with the shear-induced effects -
the \emph{shear time $\tau_S$} - can thus be defined as
\begin{equation}
\tau _{S}=\frac{1}{\Gamma_S}=\frac{1}{v q_{\|}}.
\label{eq:sheartime}
\end{equation}

The diffusion-induced (Eq.~\ref{eq:ISF_AckersonClark}) and shear-induced 
(Eq.~\ref{eq:sinx_x}) effects to the intermediate scattering functions
were shown to be independent \cite{Fuller}, hence the 
correlation functions measured in a XPCS experiment under laminar
flow can be written
\begin{multline}
g^{(2)}({\bf q},t)-1=\\
\beta\exp\left[-2 \Gamma t \left(1 - \frac{q_\perp q_\|}{q^2}\dot\gamma t 
	+ \frac{q_\|^2}{q^2}\cdot\frac{(\dot\gamma t)^2}{3}\right) \right]
      \cdot\left[\frac{\sin(\Gamma_St)}{\Gamma_St}\right]^2
\label{eq:g2flow}.
\end{multline}

The relative importance of the shear-induced effects compared to thermal 
diffusion is quantified by the ratio between the diffusion 
time (Eq.~\ref{eq:diffusiontime}), and the shear time
(Eq.~\ref{eq:sheartime}),
\begin{equation}
\text{S}=\frac{\tau_D}{\tau_S}=\frac{v q_{\|}}{D_0 q^2}
\label{eq:Snumber},
\end{equation}
which will be referred to, as the \emph{shear number}.

In order to measure the diffusion time $\tau_D$, the shear number must be
kept low (S$\ll$1). Fortunately, the shear-induced effects are visible
only in a non-transverse scattering geometry, and from Eq.~\ref{eq:Snumber} 
it results that a practical means to measure the thermal diffusion of
the scatterers is to keep the scattering wave vector perpendicular to 
the flow velocity ($q_{\|}$=0).
For any other scattering geometries, the shear number will be too high for 
almost all interesting combinations of the experimental parameters 
($\dot\gamma$, $D_0$, etc.),  
and the effect of thermal diffusion will be ``washed out'' by the much
faster decay of $g^{(2)}$ due to the shear time.  

The continuous flow of particles through the scattering volume introduces
a third relevant time scale - the transit time. Its relative importance 
compared to thermal diffusion is determined by the magnitude of the 
Deborah number, the ratio between the correlation (diffusion) time $\tau_D$ 
and the transit time, 
\begin{equation}
\text{De}=\frac{\tau_D v}{s}.
\label{eq:De}
\end{equation}
Here $v$ is the (average) flow velocity and $s$ is  
the transverse beam size (s=10$~\mu$m). 
The Deborah number is a quantity that depends not only on the flow rate and beam size,
but also on the $q$--value (through $\tau_D$), but
in all the measurements reported here, De was smaller than 0.1 even at the 
highest flow rates and smallest values of~q. Consequently, in the current
analysis, the effects induced by the finite transit time of the 
particles through the scattering volume were neglected and only the dominant 
effects related to the Brownian motion and the 
Doppler-shift induced decays of the correlation functions were considered.

In conclusion, in the limits of small De numbers (De$\ll$1) and for a 
laminar flow characterized by a single shear constant
$\dot\gamma$, we expect the correlation functions measured in an 
XPCS experiment to be described by Eq.~\ref{eq:g2flow}.
As it will be seen in the following section, the correlation functions 
measured in a transverse or a longitudinal flow geometry are well 
fitted by this form (Figs.~2~\&~3). 

\section{Results: dynamics of PMMA colloids in laminar flow}

The dynamics of the colloidal suspension of PMMA hard-spheres was probed 
in transverse and longitudinal flow geometries at average flow 
velocities ranging between 0 and 200~$\mu$m/s and for $q$-values 
in units of $qa$ ($a$ is the particle radius) 
ranging between ~3-10. At these high values of
$q$ (on scales comparable to or smaller than the particle size) 
and for the low shear rates probed here, the time-averaged static
properties are both isotropic and flow-independent (Fig.~1b~and~1c).

The dynamical properties are, however, not isotropic.
The intensity autocorrelation functions (Eq.~\ref{eq:g2flow}) can be 
written for 
the scattering geometries probed here, leading to
\begin{eqnarray}
\qperpv\ & g^{(2)}(q, t) = 1+\beta\exp\left(2\Gamma t\right)
\label{eq:g2perp}\\
\qparav\ & g^{(2)}(q, t) = 1+\beta\exp\left(2\Gamma t\right)\cdot
	\left[ \frac{\sin\left( \Gamma_S t\right)}{\Gamma_S t}\right]^2
\label{eq:g2para}.
\end{eqnarray} 
Here, the shear induced corrections to the diffusive dynamics in 
Eq.~\ref{eq:g2flow} on the order of $\dot\gamma t$ and $(\dot\gamma t)^2$
were neglected because they are too small to be measured/fitted. Also, the
shear relaxation rate $\Gamma_S$ could be related via a rheological model
(e.g. by using Eq.~\ref{eq:shearrate} and \ref{eq:sheartime}) to a shear rate
$\dot\gamma$.
This would be important if the goal was to measure the shear rate, but here
the focus is on the measurement of the diffusion dynamics (i.e.~$\Gamma$) 
hence the shear relaxation rates $\Gamma_S$ were simply obtained from the 
fits with Eq.~\ref{eq:g2para} and not related to a shear rate. 

It should also be mentioned that the X-ray contrast $\beta$ was around 
4-5~\% for all the experiments reported here, 
independent of the scattering geometry and of the flow rate. This 
agrees quite well with calculated values, 
and proves that the Deborah numbers were low enough to prevent a significant 
reduction of $\beta$ by the flow, and that the size and shape of the 
coherent beam and of the speckles were symmetric enough to prevent any
anisotropy induced by the scattering geometry.

Correlation functions measured for \qperpv\ at three different values of $q$, 
and fits with Eq.~\ref{eq:g2perp} are shown in Fig.~2. 
The correlation functions obtained at zero flow (filled symbols) are
shown together with those measured at an average flow velocity of 
$v\approx$~58.5~$\mu$m/s (empty symbols). 
The fits were
performed for all the correlation functions but for clarity, only the ones
corresponding to the $v$=58.5~$\mu$m/s data are shown (solid lines). 
As it can be seen, the correlation functions with and without flow 
are nearly identical, showing
that at these shear rates they are basically unaffected by the flow. 

\begin{figure}
\resizebox*{0.9\columnwidth}{!}{
\includegraphics{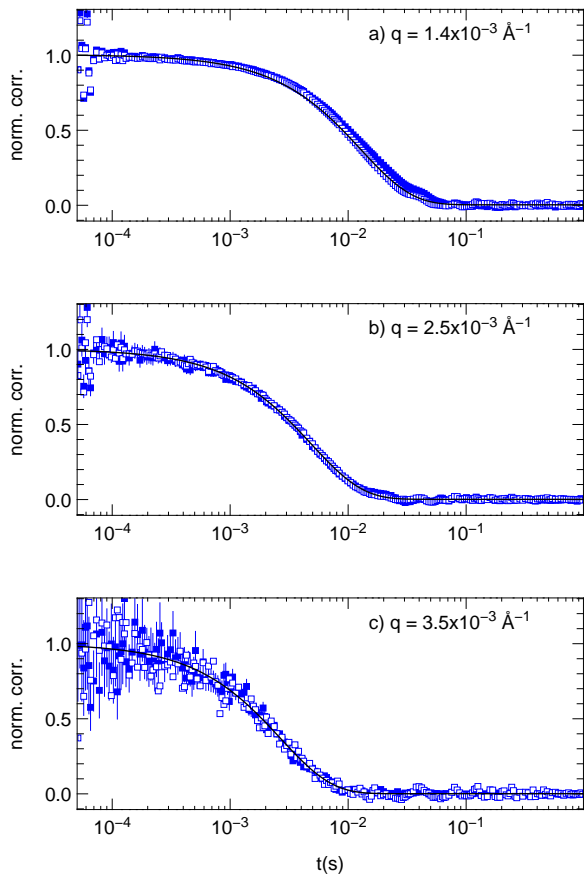}}
\caption{\label{fig:g2perp}Normalized correlation functions 
$(g^{(2)}({\bf q}, t)-1)/\beta$, obtained 
in a transverse flow geometry (\qperpv), shown here for three different
values of~$q$ at zero flow (filled symbols) and at 
$v \approx$~58.5~$\mu$m/s (empty symbols). It can be seen that for this 
flow rate, the influence of the shear flow on the correlation function is, in 
the first order, negligible. The solid lines show fits to the 
$v\approx$58.5~$\mu$m/s data with Eq.~\ref{eq:g2perp}}
\end{figure}

This conclusion is not valid for non-transverse flow geometries. 
Examples of correlation functions measured in longitudinal flow (\qparav) 
are shown in Fig.~3. Here, functions measured at a single
wave vector $q$ with a static sample ($v$=~0$\mu$m/s, panel~a) 
are shown together with the correlations measured at two
different flow rates, corresponding to average flow velocities of
$v=$11.7$\mu$m/s (b) and $v=$23.4$\mu$m/s (c). 
In the absence of flow, the correlation functions are still 
well described by simple exponential decays (Fig.~3a) allowing to obtain 
the diffusion relaxation rates $\Gamma$. These values are, within experimental
errors, comparable with the ones obtained from \qperpv -- see zero flow data 
(squares and circles) in Fig.~4. The measured diffusion
coefficient is D~$\approx $1.7$\times$10$^7$~\AA/s, with an estimated 
uncertainty of $\approx$~15\%.  

\begin{figure}
\resizebox*{0.9\columnwidth}{!}
{\includegraphics{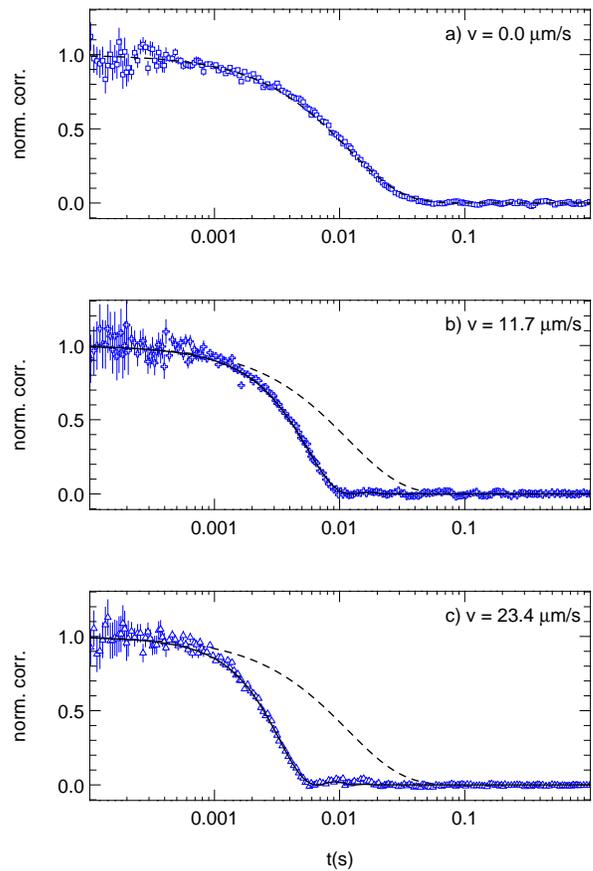}}
\caption{\label{fig:g2para}Normalized correlation functions 
$(g^{(2)}({\bf q}, t)-1)/\beta$, obtained 
in longitudinal flow scans (\qparav), at $q$=1.3$\times10^{-3}$ \AA$^{-1}$ 
and three different flow 
velocities - (a)~$v$=0, (b)~11.7~$\mu$m/s, and (c)~23.4~$\mu$m/s. The
solid lines are least square fits with Eq.~\ref{eq:g2para}. The dashed
lines show the fits to the zero flow correlation function (panel a).}
\end{figure}

The relaxations obtained from a 
flowing sample in longitudinal flow geometry  (Figs.~3b and 3c) are
strongly affected by the flow. Even though the flow velocities are smaller 
than that showed in Fig.~2 for 
\qperpv, the effects on the correlation times and on the shape of 
the correlation functions are important. 
The solid lines show the fits to the experimental data with 
Eq.~\ref{eq:g2para}. 
In principle, both relaxation rates - $\Gamma$ and $\Gamma_S$ - could be obtained from
a single fitting procedure, but due to the strong influence of
the shear-induced effects (high shear number) this is unfortunately not 
the case. 
Even at very small flow velocities (e.g. the ones used in 
Figs.~3b and 3c), the estimated 
shear numbers for a longitudinal scattering wave vector 
$q_\|=$~1.3$\times$10$^{-3}$\AA$^{-1}$ are $S \approx$ 3.6 (b) and 
$S \approx$ 7.2 (c). 
As a consequence, the intensity correlation functions
are dominated by the shear time if a non-transversal
flow geometry is used, 
and the errors on the fitted values for $\Gamma$ are high. 

It should also be mentioned that the oscillations which can be noticed on 
some of the correlation functions at long times (Figs.~2~\&~3) are not due 
to the shear effects described by Eq. 13. As they tend to appear/disappear 
on a more ``random'' basis, a possible explanation would be that bubbles 
or other impurities, sometimes sweep through the scattering volume. 
As the statistical error bars are smaller at longer times  
such artifacts may appear like real effects.

In order to measure the diffusive dynamics of the particles under flow, 
the correlation times (relaxation rates) must be obtained from
the \qperpv\ data.  
The dispersion relationships for the diffusion coefficient 
$D=\Gamma q^{-2}$ measured in transverse flow scans
at zero flow as well as two (relatively) high flow velocities 
$v$=58.5~$\mu$m/s and $v$=117~$\mu$m/s 
are shown in Fig.~4. As it is expected for
a suspension undergoing Brownian motion, the
correlation times measured at $v$=0 and $v$=58.5~$\mu$m/s 
decay as $q^{-2}$ and the diffusion coefficient is
$q$-independent. In addition, one can observe that in this scattering 
geometry, the measured relaxation rates (and the diffusion coefficient)
are also flow-independent. This conclusion holds to a certain degree for the
$v$=117~$\mu$m/s data as well, although at lowest values of~$q$ the measured 
diffusion coefficients start being enhanced by the shear.

\begin{figure}
\resizebox*{0.99\columnwidth}{!}{
\includegraphics{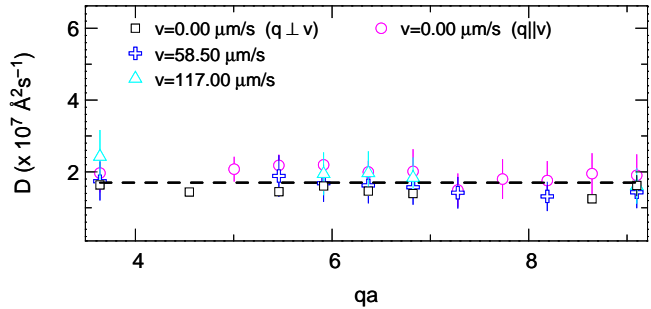}} \caption{\label{fig:tauq2}
Diffusion constant $D$ associated with the dynamics of the 
colloidal particles (radius $a$) in shear flow as a function of $q$ 
measured in a transverse flow (\qperpv) geometry for three
volume flow rates (squares, crosses and triangles). Values for the diffusion
coefficient obtained from longitudinal flow scans (\qparav) with a
static samples are also shown (circles).
The solid line shows the estimated value for the 
diffusion constant (see text).}
\end{figure}

The diffusion coefficient can also be estimated using the 
Einstein-Stokes relationship,
\begin{equation}
D_0=\frac{k_B T}{6 \pi \eta a}.
\label{eq:StokesEinstein}
\end{equation}
The viscosity of the solvent is $\eta_0\approx$3.0$\times$10$^{-3}$~Pa$\cdot$s,
and the viscosity of the solution is approximately \cite{Segre_PRL95} 
$\eta\approx$1.5$\cdot \eta_0$ = 4.5$\times$10$^{-3}$~Pa$\cdot$s.
With values corresponding to our experimental conditions
(T$\approx$295K and $a\approx$255 nm), the resulting diffusion
coefficient is $D_0\approx$1.88$\times$10$^7$~\AA$^2$/s, which is in good
agreement with the values measured by XPCS 
(D$\approx$~1.7$\times$10$^7$~\AA/s, dashed line in Fig.~4). The small discrepancy could be attributed to a slightly different viscosity of the 
solution and/or to small hydrodynamic effects.

\section{Conclusions}

The method presented here allows the direct measurement of mesoscale
dynamics in a complex fluid under laminar flow, and can be used 
to obtain diffusion coefficients and/or to retrieve of information 
about particle size, rheological properties of the solvent, etc. 
in a variety of X-ray sensitive samples. 

The flow-induced transit-time effects are negligible
for most of the interesting combinations of sample and flow properties except 
perhaps the most viscous suspensions \cite{Busch_EPJE}, but the
measured correlation functions are, in most situations, strongly affected by 
the shear time. The basic idea of the method presented here,
is to keep the shear-induced effects on the correlation functions as low 
as possible by choosing a transverse flow geometry (\qperpv) and a 
low-enough shear rate. 

For a perfect transverse flow 
alignment, $q_{\|}$=0 and the shear time (Eq.~\ref{eq:sheartime}) is 
infinity. As a consequence 
the shear number is zero and a XPCS experiment measures only the 
diffusion time. In practice, such a perfect alignment does not exist 
and a more realistic shear number (Eq.~\ref{eq:Snumber}) can be written 
as
\begin{equation}
S=\frac{v\phi}{D_0 q},
\label{eq:SPhinumber}
\end{equation}
where $\phi$ is a small misalignment angle (in radians) of the flow channel 
with respect to the longitudinal direction. 

Setting an upper limit on the shear number 
(e.g. S$\approx$0.1) in order to keep the shear-induced effect low, 
leads to a maximum acceptable value for the shear rate $\dot\gamma$ 
\cite{Salmon}. This value is dependent on the exact nature of the 
sample ($D_0$), on the scattering wave vector $q$, and on the 
particular alignment (angle~$\phi$).

\begin{figure}
\resizebox*{0.9\columnwidth}{!}{
\includegraphics{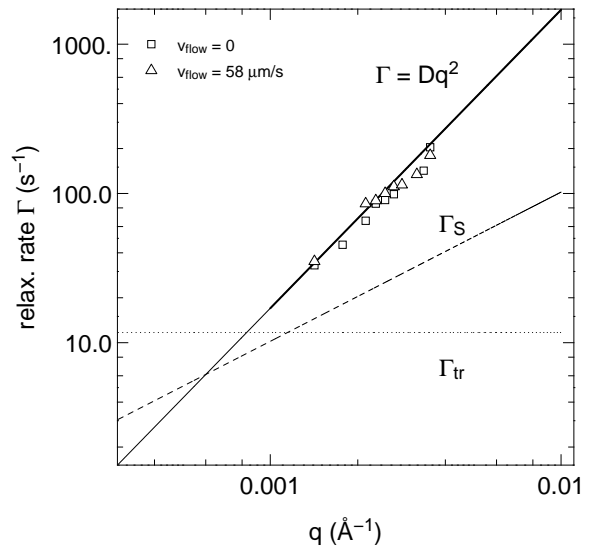}} \caption{\label{fig:DifShTr}
Dispersion relationships for the diffusion ($\Gamma = D q^2$), 
shear ($\Gamma _S= v \phi q$), and transit ($\Gamma_{tr}=v/s$) relaxation 
rates. The (example) values used here to estimate $\Gamma$, $\Gamma _S$, 
and $\Gamma _{tr}$ were, $D \approx$~1.7$\times$10$^7$~\AA/s, $\phi\approx$
0.01 (0.5~deg), $v=58.5 \mu$m/s. The thick solid line highlights the~$q$ region 
that could, in principle, be accessed by XPCS. With the flow parameters
chosen here, this is also the region where the thermal diffusion dominates 
the dynamic signal and can be measured by XPCS.
The experimental points show the relaxation rates measured with the static, 
$v=0$ (squares), and the $v=58.5~\mu$m/s (triangles),
samples.}
\end{figure}

The experimental conditions under which it is possible to measure the
diffusive dynamics of the particles can be ``visualized'' 
in Fig.~5. Here, the dispersion relationships for the
diffusion rate $\Gamma$  and the shear relaxation rate $\Gamma _S=v\phi q$
(assuming a ``nearly transverse'' flow geometry with a small misalignment
angle $\phi$) were plotted together with the ($q$-independent) transit rate. 
As stated above, the shear and Deborah numbers must be much smaller 
than unity, or equivalently the diffusion relaxation rate must be 
much higher than the shear- and transit- induced relaxation rates. 
With the values chosen as an example in Fig.~5, 
$\phi$=0.01 (corresponding to an assumed misalignment of $\approx$0.6~deg),
D = 1.8$\times$10$^7$~\AA$^2$/s, and a flow velocity of $v$=~58.5$\mu$m/s,
this condition is fulfilled for the $q$-range accessible 
by XPCS (highlighted by the thick solid line), and the relaxation rates
measured at $v$=~0 and 58.5~$\mu$m/s confirm that this conditions are 
fulfilled. At the same time, it is clear that this would not be the case if 
the flow velocity increased beyond a maximum acceptable value, or if the
measurements were performed at smaller $q$, for instance by DLS. The maximum 
flow velocity that allows measurements of the diffusive dynamics depends on the
sample (time scales that have to be measured), the flow geometry and the
scattering alignment. The study presented here is a proof of principle,
and the data show that the transverse flow alignment was achieved with an 
accuracy better than $\approx$0.5~deg, and that flow velocities up to 
$\approx$50-100~$\mu$m/s allow measurements of the diffusive dynamics 
of the colloidal suspension. In principle such a value (e.g. for the maximum
acceptable flow velocity) could be used as a reference, and it should be
possible to scale it in order to determine the experimental conditions that
allow measurements of the diffusive dynamics on different samples, with
different relaxation times. These ideas will be explored in further studies.

The method described here allows also measurements of various properties 
of the laminar flow (e.g. using the shear relaxation rates $\Gamma_S$ 
measured in a longitudinal flow geometry to calculate the
shear rate $\dot\gamma$). While this is a valid aim  
\cite{Fuller,Narayan_AO97}, it was not our purpose here. 
We intended to demonstrate that the diffusive dynamics is accessible 
in a flowing sample. 
We believe that measuring the ``intrinsic'' dynamical properties of the
fluid sample in a microfluidic experiment/setup, 
provides a tremendous amount of interesting opportunities for XPCS 
experiments in soft-matter and biological systems. The minimal requirements 
for such experiments would be a strong enough scattering from the sample - 
say,~$\approx$1~photon per speckle per correlation time, if a 2D area detector is used,
and a flow/shear rate which is fast enough to prevent beam damage but small
enough to allow the measurement of diffusive dynamics in the
presence of shear.

We wish to acknowledge helpful discussions with Narayanan Theyencheri,
Jean-Baptiste Salmon, Fanny Destremaut, Yuriy Chushkin, Sebastian Busch, 
Erik Geissler and Mark Sutton.



\bibliography{references}
\end{document}